\documentclass[prl,aps,superscriptaddress,twocolumn,showpacs,preprintnumbers,nofootinbib]{revtex4-1}
\usepackage{epsfig}
\newcommand{\usb}{\affiliation{Departamento de F\'{\i}sica, Secci\'{o}n de Fen\'{o}menos \'{O}pticos, 
Universidad Sim\'{o}n Bol\'{\i}var,Apartado Postal 89000, Caracas 1080-A, Venezuela.}}
\newcommand{\ivic}{\affiliation{Centro de F\'{\i}sica, Instituto Venezolano de Investigaciones 
Cient\'{\i}ficas, Apartado 20632 Caracas 1020-A, Venezuela.}}

\begin{document}
\begin{flushright} ${}$\\[-40pt] $\scriptstyle \mathrm SB/F/478-18$ \\[0pt]
\end{flushright}
\title{Momentum exchange  between an electromagnetic wave and a dispersive medium}
\author{Rodrigo Medina}\email[]{rmedina@ivic.gob.ve}\ivic
\author{J Stephany}\email[]{stephany@usb.ve}\usb
\date{\today}

\begin{abstract}
We present an elementary discussion of the momentum transferred  by an electromagnetic wave propagating 
in a dispersive medium.  Our analysis is based on Minkowski's electromagnetic momentum density 
which have been recently seen to be consistent with a fully covariant expression of the energy-momentum 
tensor of the electromagnetic field in a dispersive medium and with all the experimental evidence. 
We show that  the medium may be either pulled or pushed as the electromagnetic 
pulse enters in it depending on the value of the frequency.

\end{abstract}
\maketitle
The description of momentum exchange between an electromagnetic wave and a dielectric media is
an intriguing issue which has been dominated historically by Abraham-Minkowski's controversy.
Initially the controversy was focused on which, between Minkowski's density 
\begin{equation}
\label{Minmomentum-density}
\mathbf{g}_{\mathrm{Min}} = c^{-1} T^{i0}_\mathrm{Min} \hat{\mathbf{e}}_i =
\frac{1}{4\pi c}\mathbf{D}\times\mathbf{B}\ 
\end{equation}
and Abraham density 
\begin{equation}
\label{abrmomentum}
\mathbf{g}_{\mathrm{Abr}}=c^{-2}\mathbf{S}=\frac{1}{4\pi c}\mathbf{E}\times\mathbf{H}\ 
\end{equation}
should be taken as the momentum density of the electromagnetic field. Later on, it was recognized the necessity of
identifying the electromagnetic contributions to the matter energy-momentum tensor and the importance of a better
understanding of the criteria which allow to distinguish between matter and field. 
In the process, it became necessary to   
re-interpret even such a familiar concept as Poynting's energy density.  Poynting \cite{Poynting} proposed his energy 
conservation equation for homogeneous isotropic materials with time-independent linear susceptibilities 
 $\mathbf{D}=\epsilon\mathbf{E}$ and $\mathbf{H}=1/\mu\mathbf{B}$ in 1884.  He proposed to generalize  the 
energy density of the field in vacuum $u_{\mathrm{vac}} = \frac{1}{8\pi}(E^2+B^2)$ to
\begin{equation}
\label{Poynting-energy}
u_\mathrm{P}=\frac{1}{8\pi}(\mathbf{E}\cdot\mathbf{D}+
\mathbf{H}\cdot\mathbf{B}) \ .
\end{equation}
With the mentioned restrictions on the medium it follows from
\begin{equation}
\label{energy-deriv}
\frac{\partial u_\mathrm{P}}{\partial t}=\frac{1}{4\pi}\Big(\mathbf{E}\cdot
\frac{\partial \mathbf{D}}{\partial t}+
\mathbf{H}\cdot\frac{\partial \mathbf{B}}{\partial t}\Big) 
\end{equation}
and Maxwell equations that
\begin{equation}
\label{Poynting-eq}
 \frac{\partial u_\mathrm{P}}{\partial t} +\nabla\cdot\mathbf{S} =
-\mathbf{E}\cdot\mathbf{j}_{\mathrm f}\ ,
\end{equation}
where Poynting's vector $\mathbf{S}$ was defined above.
In this view $\mathbf{S}$, is the energy current density of the
electromagnetic disturbance, and $\mathbf{E}\cdot\mathbf{j}_{\mathrm f}$ is the time 
rate of work done by the field on free charges. No work is done on the polarizable 
matter. Poynting's construction is not valid for non linear polarizable matter. In our recent 
work \cite {MedandSb,MedandSg} we have shown that (\ref{Poynting-energy}) does not represent 
the field energy alone but corresponds to the energy of the whole electromagnetic wave,  understood 
as a mixed entity with contributions of the field and the polarizations. We first identify
the force density on matter to be given by,
\begin{equation}
 \label{dipolarforce}
 f^\mu_\mathrm{d}=\frac{1}{2}D_{\alpha\beta}\partial^\mu F^{\alpha\beta}\ ,
\end{equation}
with  $D_{\alpha\beta}$ the  space-time dipolar density, whose  spatial part is the magnetization density
$D_{ij}=\epsilon_{ijk}M_k$ and whose temporal part is the polarization $D_{0k}=-D_{k0}=P_k$. Then we
show that the energy momentum tensor of the electromagnetic field compatible with this force  is given by
\begin{equation}
\label{energy-momentum-tensor}
T^{\mu\nu}_{\mathrm{F}}=
-\frac{1}{16\pi}F_{\alpha\beta}F^{\alpha\beta}\eta^{\mu\nu}
+\frac{1}{4\pi}F^{\mu}_{\ \alpha}H^{\nu\alpha}\ ,
\end{equation}
where $H^{\mu\nu}=F^{\mu\nu}-4\pi D^{\mu\nu}$ and $\eta^{\mu\nu}$ is the metric tensor
with the signature $(-,+,+,+)$. 
For this tensor, which is valid in particular for a 
dispersive polarizable medium, the energy density is
\begin{equation}
\label{energy-density}
u_{\mathrm{F}} = T_{\mathrm F}^{00} = \frac{1}{8\pi}(E^2+B^2)+\mathbf{E}\cdot\mathbf{P} \ ,
\end{equation}
the energy current density $ c T_{\mathrm F}^{0i}=\mathbf{S} =c\mathbf{E}\times
\mathbf{H}/4\pi$ is Poynting's vector and the momentum density $ c^{-1}T_{\mathrm F}^{i0}
=\mathbf{g}_{\mathrm{F}} =\mathbf{D}\times
\mathbf{B}/4\pi c$ coincides with Minkowski's expression  (\ref{Minmomentum-density}).  
Maxwell's stress tensor is given by,
\begin{eqnarray}
\label{Maxwell-tensor}
 T_{\mathrm F}^{ij}  = \frac{1}{8\pi}(E^2&+&B^2)\delta_{ij} 
-\mathbf{B}\cdot\mathbf{M}\delta_{ij} \nonumber\\
 &-&\frac{1}{4\pi}(E_i D_j +H_i B_j) \ .
\end{eqnarray}
The difference between the energy density above  and the energy density of the
vacuum, $u_\mathrm{F}-u_\mathrm{vac}=\mathbf{P}\cdot\mathbf{E}$, is the negative of the
electrostatic energy density of polarization which should be considered part of the energy of
matter. This makes physical sense because it contributes to the inertia of matter in 
exactly the same way  as nuclear interaction energy contributes to nuclei
mass.  Note that there is no similar magnetic term since no potential
magnetic energy exists. The energy density of matter is given by,
\begin{equation}
\label{matter-energy-density}
u_\mathrm{M}(\mathbf{P},\mathbf{M},\mathbf{E})=
u_\mathrm{b}(\mathbf{P},\mathbf{M})-\mathbf{P}\cdot\mathbf{E}
\end{equation}
where $u_\mathrm{b}$ is the bare energy density that does not depend on the electromagnetic 
field. It may be splitted in a term which does not depend on $\mathbf{P}$ and $\mathbf{M}$ 
and one that does,
\begin{equation}
u_\mathrm{b}(\mathbf{P},\mathbf{M})=u_\mathrm{b0}+
u_\mathrm{PM}(\mathbf{P},\mathbf{M})
\end{equation}
Similarly the momentum density of 
matter is shown to have the structure,
\begin{equation}
\label{matter-momentum-density}
\mathbf{g}_\mathrm{M}(\mathbf{P},\mathbf{M},\mathbf{E},\mathbf{B}) = 
\mathbf{g}_\mathrm{b}(\mathbf{P},\mathbf{M},\mathbf{E})
-\frac{1}{c}\mathbf{P}\times\mathbf{B}\ .
\end{equation}
When matter is immersed in an electromagnetic field the energy density
of matter changes. This energy difference can be calculated integrating the 
power expression obtained from (\ref{dipolarforce}). For time-independent linear polarizabilities 
the work done on matter does not depend on the way the fields change in time, it depends only on 
the final values of the fields
\begin{eqnarray}
\Delta u_\mathrm{M} &=& u_\mathrm{M}(\mathbf{E},\mathbf{B})- u_\mathrm{M}(0,0)=
\int dw_\mathrm{d}\nonumber\\
&=& -\frac{1}{2}(\mathbf{E}\cdot\mathbf{P}+\mathbf{B}\cdot\mathbf{M}) \ .
\end{eqnarray}
Since it includes the electrical potential energy
density $- \mathbf{E}\cdot\mathbf{P}$ we can write
\begin{equation}
\Delta u_\mathrm{M}=-\mathbf{E}\cdot\mathbf{P}+
\frac{1}{2}(\mathbf{E}\cdot\mathbf{P}-\mathbf{B}\cdot\mathbf{M})\ .
\end{equation}
Poynting's energy density turns out to be 
\begin{equation}
u_\mathrm{P}=u_{\mathrm{F}}+\Delta u_\mathrm{M}=\frac{1}{8\pi}(E^2+B^2)+
\frac{1}{2}(\mathbf{E}\cdot\mathbf{P}-\mathbf{B}\cdot\mathbf{M})\ .
\end{equation}
supporting the interpretation mentioned above. In \cite{MedandSg}, we discuss how
this approach is consistent with all the experimental evidence

When these ideas are applied to the elementary example of an electromagnetic wave 
propagating in a linear homogeneous polarizable medium interacting with a conducting sheet 
\cite{MedandSf} we found that conservation of momentum is achieved when Minkowski's expression,
which is consistent with our approach, is used but
not when Abraham's expression is considered.

In this letter  we are interested in momentum exchange between an electromagnetic pulse  and a dispersive medium
in the simplest case, when only a resonant frequency appears in the dispersion relation. 
As the electromagnetic wave passes through the medium it induces a dipole momentum in the atoms but in
first approximation, it does not produce any appreciable changes in its positions. In a simple textbook approach
we can  take the medium as a collection of  damped harmonic oscillators satisfying,
\begin{equation}
 m\ddot\mathbf{r}=-k\mathbf{r}-{\alpha m}\dot\mathbf{r}-e\mathbf{E}
\end{equation}
with $m$,$k$ and $\alpha$ phenomenological constants. The dipolar momentum of the atom is $\mathbf{p}=-e\mathbf{r}$ 
and satisfies,
\begin{equation}
 \ddot\mathbf{p}=-\omega_0^2\mathbf{p}-\alpha\dot\mathbf{p}+\frac{e^2}{m}\mathbf{E}
\end{equation}
In an homogeneous material the polarization $\mathbf{P}$ is linear in the individual moments and we can 
write
\begin{equation}
\label{EqPol}
\ddot\mathbf{P}=-\omega_0^2\mathbf{P}-\alpha\dot\mathbf{P}+\chi_0\omega_0^2\mathbf{E}
 \end{equation}
with $\chi_0$ a constant with  susceptibility units. Independently of the simple model from which it 
has been deduced we take Eq.(\ref{EqPol}) as the one which is characteristic of a dispersive medium with a single
resonance. Note that the same equation is obtained 
up to second order in the quantum mechanically perturbative treatment.

For a plane wave in the usual complex notation
we have the fields
\begin{equation}
 \hat\mathbf{E}=\mathbf{E}e^{-i\omega t}\ \ \ ,\ \ \  \hat\mathbf{P}=\mathbf{P}e^{-i\omega t}\ .
\end{equation}
related by the susceptibility $\chi (\omega)$ by,
\begin{equation}
 \mathbf{P}=\chi (\omega)\mathbf{E}\ .
\end{equation}
Substituting in the equation we have the usual single resonant expression for the susceptibility
\cite{Marion1980}
\begin{equation}
 \chi(\omega)=\frac{\chi_0\omega_0^2}{\omega_0^2-\omega^2-i\alpha\omega}
\end{equation}
Let us first discuss energy exchange. Multiplying the equation above by $\dot\mathbf{P}$ we have,
\begin{equation}
 \frac{1}{2}\frac{d}{dt}\dot\mathbf{P}^2+\frac{\omega_0^2}{2}\frac{d}{dt}\mathbf{P}^2
 -\chi_0\omega_0^2\frac{d}{dt}(\mathbf{E}\cdot\mathbf{P})+\chi_0\omega_0^2\dot\mathbf{E}\cdot\mathbf{P}
 =-\alpha\dot\mathbf{P}^2
 \end{equation}
This is expressed more conveniently as,
\begin{equation}
\label{powerbalance}
 \frac{d}{dt}\left( \frac{1}{2\chi_0\omega_0^2}\dot\mathbf{P}^2+\frac{1}{2\chi_0}\mathbf{P}^2-
 \mathbf{E}\cdot\mathbf{P}\right)=
-\frac{\alpha}{\chi_0\omega_0^2}\dot\mathbf{P}^2-\dot\mathbf{E}\cdot\mathbf{P}
\end{equation}
The first term in the right hand side corresponds to the energy loses of our model due to damping
and goes to increase  $u_\mathrm{b0}$. 
According to the  relativistic expression of the force density  (\ref{dipolarforce}),
the second term is the power which the field gives to matter
\begin{equation}
\label{power-density-dip}
cf^0_\mathrm{d}= -\mathbf{P}\cdot\frac{\partial\mathbf{E}}{\partial t}-
\mathbf{M}\cdot\frac{\partial\mathbf{B}}{\partial t}\ .
\end{equation}
Energy equilibrium is achieved  by taking
\begin{equation}
 u_\mathrm{PM}(\mathbf{P},\mathbf{M})=\frac{1}{2\chi_0\omega_0^2}\dot\mathbf{P}^2+\frac{1}{2\chi_0}\mathbf{P}^2
\end{equation}
in our model. The first term in (\ref{powerbalance}) is naturally 
interpreted as the kinetic energy and the second term as the elastic term. We insist that
$-\mathbf{E}\cdot\mathbf{P}$ is the electrostatic polarization energy.

So, according to (\ref{matter-energy-density}) and 
(\ref{energy-density}) we have  the expressions,
\begin{eqnarray}
 u_\mathrm{M}&=&u_\mathrm{b0}+\frac{1}{2\chi_0\omega_0^2}\dot\mathbf{P}^2+\frac{1}{2\chi_0}\mathbf{P}^2-
 \mathbf{E}\cdot\mathbf{P}\\
 u_\mathrm{F}&=&\frac{1}{8\pi}(E^2+B^2)+\mathbf{E}\cdot\mathbf{P}
\end{eqnarray}
for the matter and field energy densities. Here $ u_\mathrm{b0}$ is the non electromagnetic
energy density of matter. The energy of the full electromagnetic wave is,
\begin{eqnarray}
\label{uwave}
 u_\mathrm{W}&=& u_\mathrm{F}+u_\mathrm{M}-u_\mathrm{b0}\nonumber\\
 &=&\frac{1}{8\pi}(E^2+B^2)+\frac{1}{2\chi_0\omega_0^2}\dot\mathbf{P}^2+\frac{1}{2\chi_0}\mathbf{P}^2
\end{eqnarray}
This generalizes Poynting's formula which is recovered in the static case.

\begin{figure}
\centering
\includegraphics[scale=0.5]{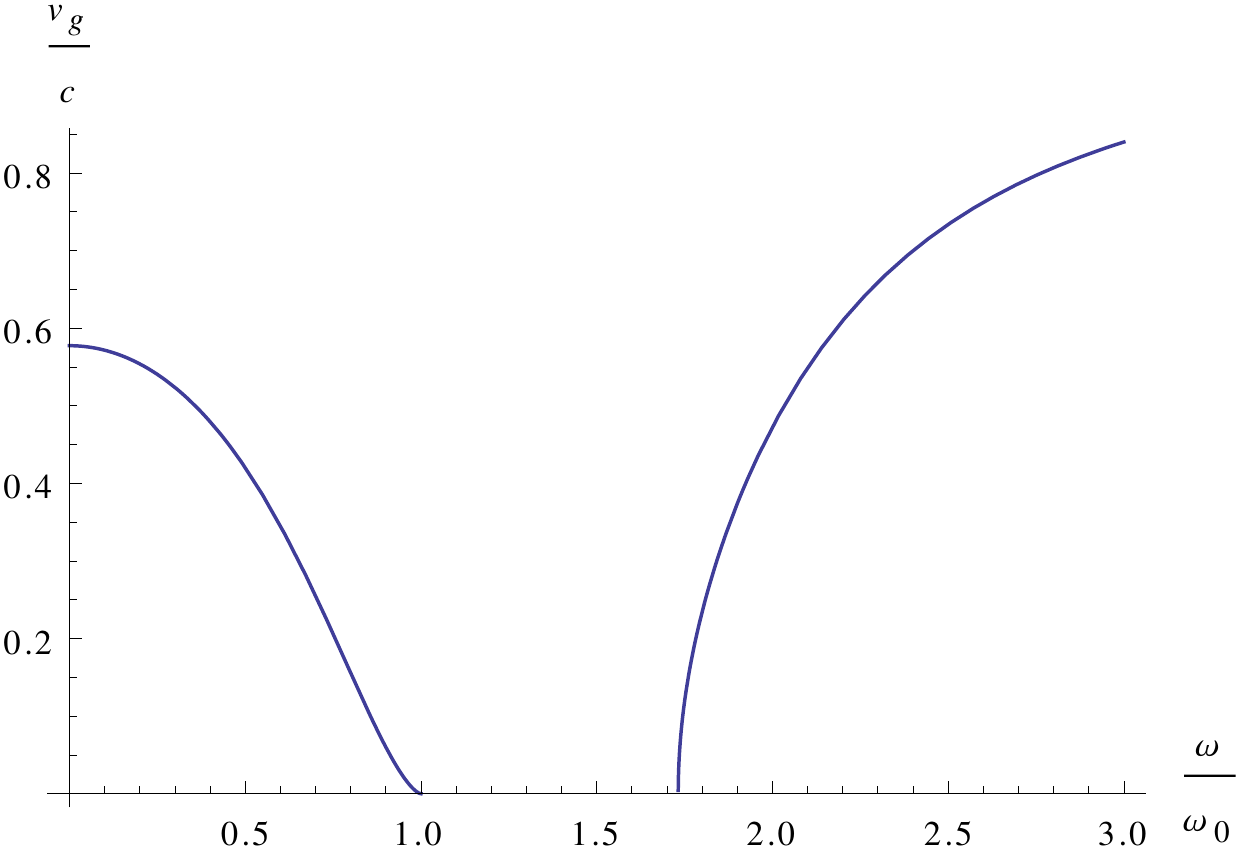}
\caption{The normalized group velocity $v_g/c$}
\label{vg/c}
\end{figure}

In order to test this result in a  specific situation we consider a dilute gas of
polar atoms. We use the simplified textbook description of the phenomenon,
assuming that there is no appreciable interaction between the atoms that constitute the gas
and that the losses are negligible, $\alpha\approx 0$ . We 
suppose that the gas is confined in some
region but that the boundary is diffuse so that there is not a reflected ray when the light hits the gas. 
We consider a wave packet coming from vacuum with a transversal section $A$ and a length $cT$  with $T$ some time interval. 
For definiteness we take the electromagnetic pulse traveling in the direction of $\hat\mathbf{e}_1$. 
It has an energy density,
\begin{equation}
 u=\frac{1}{8\pi}(E^2+B^2)=\frac{1}{4\pi}(E^2)
\end{equation}
and a time averaged energy density,
\begin{equation}
 \langle u\rangle=\frac{1}{8\pi}(E_0^2)
\end{equation}
which results in a total energy in vacuum given by
\begin{equation}
 U_0=cTA\frac{E_0}{8\pi}
\end{equation}
Here $E_0$ is the amplitude of the electric field in vacuum. In the medium the wave takes the form \cite{Marion1980},
\begin{equation}
 \mathbf{E}={E}_1e^{i(kx_1-\omega t)}\hat\mathbf{e}_2\ \ ,\ \  \mathbf{B}={B}_1e^{i(kx_1-\omega t)}\hat\mathbf{e}_3
\end{equation}
where using $\epsilon=1+4\pi\chi(\omega)$, we have
\begin{equation}
k=\frac{\omega}{c}\sqrt{\epsilon}=\frac{\omega}{c}\sqrt{1+4\pi\chi(\omega)} \ \ ,\ \ B_1=\sqrt{\epsilon}E_1\ .
\end{equation}
It has a phase velocity given by 
\begin{equation}
 v_{ph}=\frac{\omega}{k}=\frac{c}{\sqrt{1+4\pi\chi(\omega)}}
\end{equation}
When the denominator is negative the medium is opaque. The group velocity is computed from
\begin{equation}
 v_g=\frac{d\omega}{dk}=\frac{1}{\frac{dk}{d\omega}}\ .
\end{equation}
Doing the algebra one gets,
\begin{equation}
 \frac{dk}{d\omega}=\frac{1}{v_{ph}(1+4\pi\chi(\omega))}[1+4\pi\chi(\omega)\frac{\omega_0^2}{\omega_o^2-\omega^2}]
\end{equation}
\begin{equation}
 v_g=\frac{v_{ph}(1+4\pi\chi(\omega))}{1+4\pi\chi(\omega)\frac{\omega_0^2}{\omega_o^2-\omega^2}}
 =c\frac{\sqrt{(1+4\pi\chi(\omega))}}{[1+4\pi\chi(\omega)\frac{\omega_0^2}{\omega_0^2-\omega^2}]}
\end{equation}
The group velocity is always smaller than $c$. In Figure \ref{vg/c}, we show the dependency of
the group velocity with the frequency. At low frequencies $v_g\longrightarrow v_{ph}$.
\label{vgentrec}

After the pulse  has completely entered the medium it has a length $v_g T$ and we will suppose that the energy loses are 
depreciable. The energy of the wave in the gas should be computed using (\ref{uwave}). For the time average 
we have to take care of the phase difference between $\mathbf{P}$ and $\dot\mathbf{P}$. Denoting by $E_1$ and $P_1$
the magnitudes of $\mathbf{E}$ and $\mathbf{P}$ in the medium, we have $P_1=\chi(\omega)E_1$ and we get,
\begin{equation}
 \langle \frac{1}{2\chi_0\omega_0^2}\dot\mathbf{P}^2+\frac{1}{2\chi_0}\mathbf{P}^2 \rangle=
 \frac{P_1^2(\omega_0^2+\omega^2)}{4\chi_0\omega_0^2}
\end{equation}
Then,
\begin{eqnarray}
 \langle u_W\rangle
 &=&\frac{1}{16\pi}(E_1^2+\frac{c^2}{v_{ph}^2} E_1^2)+
 \frac{\chi^2(\omega)E_1^2(\omega_0^2+\omega^2)}{4\chi_0\omega_0^2}\nonumber\\
 &=&\frac{E_1^2}{8\pi}[1+4\pi\chi(\omega)\frac{\omega_0^2}{\omega_0^2-\omega^2}]
\end{eqnarray}
Multiplying by the volume we get the total energy,
\begin{equation}
 U_1=\langle u_W\rangle Av_gT=\frac{E_1^2}{8\pi}AcT\sqrt{1+4\pi\chi(\omega)}
\end{equation}
Since we are supposing that energy is conserved $U_0=U_1$, we obtain
the relation between the amplitudes,
\begin{equation}
 E_1^2=\frac{E_0^2}{\sqrt{1+4\pi\chi(\omega)}}=\frac{E_0^2}{\sqrt{\epsilon}}
\end{equation}
The momentum density in vacuum is
\begin{equation}
 \mathbf{g}_0=\frac{1}{4\pi c}\mathbf{E}\times\mathbf{B}=\frac{1}{4\pi c}E^2\hat\mathbf{e}_1
\end{equation}
with the time average,
\begin{equation}
 \langle \mathbf{g}_0\rangle=\frac{1}{8\pi c}E_0^2\hat\mathbf{e}_1\ .
\end{equation}
\begin{figure}
\centering
\includegraphics[scale=0.5]{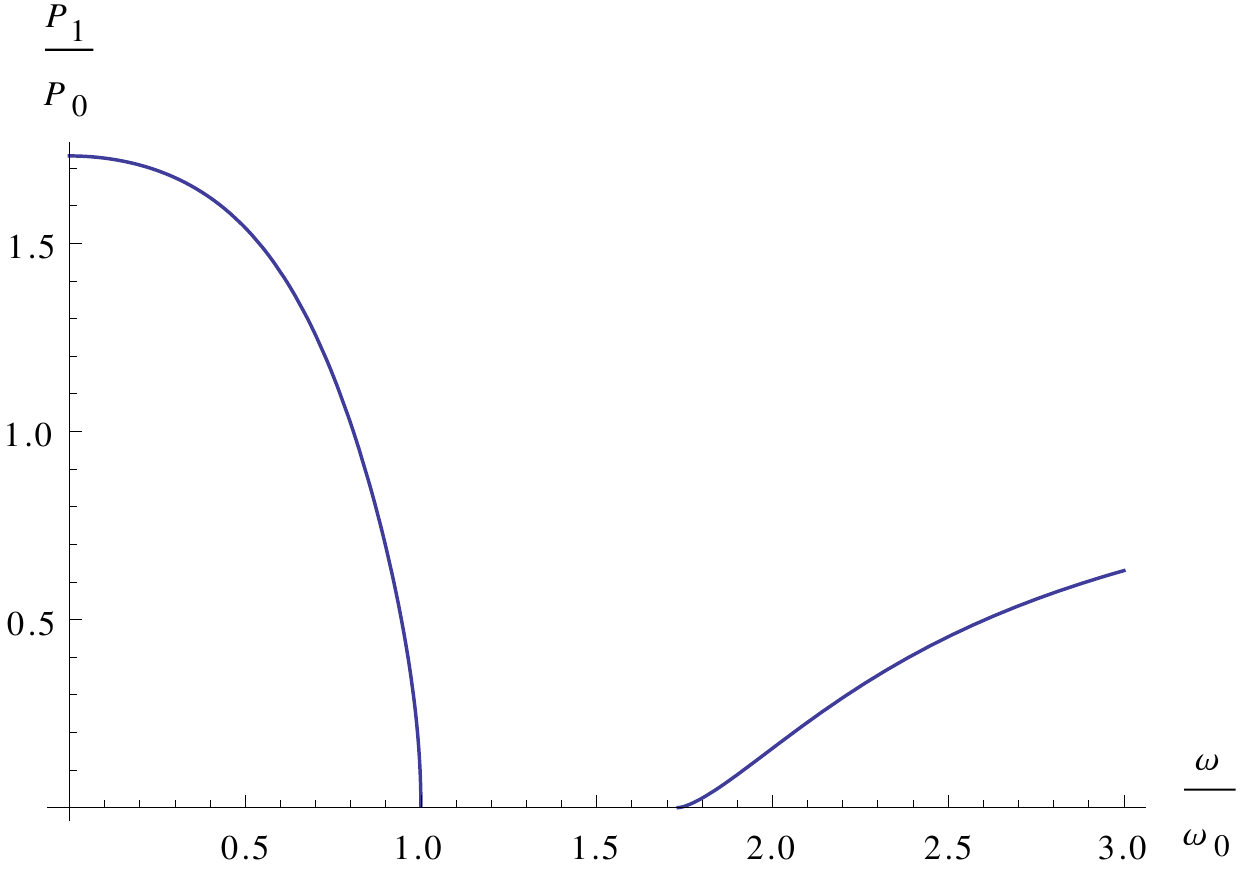}
\caption{The transmitted vs incoming  momentum ratio $P_1/P_0$}
\label{P1/P0}
\end{figure}
 The total momentum in vacuum is,
 \begin{equation}
  \mathbf{P}_0=\langle \mathbf{g}_0\rangle AcT=\frac{1}{8\pi}E_0^2AT\hat\mathbf{e}_1
 \end{equation}
 In the medium, using Minkowski's density
 \begin{equation}
   \mathbf{g}_0=\frac{1}{4\pi c}\mathbf{D}\times\mathbf{B}=\frac{1}{4\pi c}DB\hat\mathbf{e}_1
\end{equation}
and
\begin{equation}
 \langle \mathbf{g}_1\rangle=\frac{1}{8\pi c}D_1B_1\hat\mathbf{e}_1=
 \frac{(1+4\pi\chi(\omega))}{8\pi v_{ph}}E_1^2 \hat\mathbf{e}_1.
 \end{equation}
 \begin{equation}
 \langle \mathbf{g}_1\rangle=
 \frac{\sqrt{(1+4\pi\chi(\omega))^3}}{8\pi c}E_1^2\hat\mathbf{e}_1=\frac{\epsilon E_0^2}{8\pi c}\hat\mathbf{e}_1
 .
 \end{equation}
Finally,
\begin{equation}
 \mathbf{P}_1=\langle \mathbf{g}_1\rangle Av_gT=
 \frac{\sqrt{(1+4\pi\chi(\omega))^3}}{[1+4\pi\chi(\omega)\frac{\omega_0^2}{\omega_0^2-\omega^2}]}\mathbf{P}_0
\end{equation}
\begin{equation}
 \mathbf{P}_1=
 \epsilon\frac{v_g}{c}\mathbf{P}_0
\end{equation}
Figure \ref{P1/P0} displays the behavior of the transmitted electromagnetic momentum. Below the 
resonance frequency it is larger than the incoming momentum. Conservation of momentum  
implies that the gas acquires mechanical momentum in the direction opposite to the propagation of the 
wave. As the frequency  increases the excess of transmitted electromagnetic momentum diminishes. 
At some value the transmission is recoilless.  Finally near the resonant frequency the transmitted 
electromagnetic momentum is only a fraction of  the incoming momentum  and the gas acquires mechanical 
momentum in the direction of  propagation of the wave. For frequencies above the opaque region the gas 
also acquires mechanical momentum in the direction of propagation of the wave.

The simplicity of the picture presented in this paper suggests that it may be also useful to describe
the main features of momentum transmission in other more complicate situations. 
In an  example discussed recently
\cite{Khamehchi2017}, the anisotropic expansion of a Bose-Einstein condensate when one of the light beams of 
the confining traps is switched off was portrayed as a negative effective mass effect. Instead, 
it can be view as the result of  momentum transfer by the  transmitted beam. We note again that the 
main features of our computation translate to the perturbative quantum mechanical treatment. As another more 
familiar situation where the work presented in this paper may be of  relevance, we may point to 
the experiments discussed in \cite{AshkinD1973,Casner2001}. There, it was shown that light  either entering 
or leaving a steady liquid exerts a net outward force at the liquid surface. Nevertheless for the discussion 
of this experiments one should include the reflected wave and the possibility of multiple resonant frequencies.
A related calculation  was presented in  \cite{MedandSb} where we showed, 
using the  solution of Maxwell's 
equations, that an incoming  wave which enters a region filled by an homogeneous dielectric
medium exerts on it a force which is opposite to the direction in which it propagates.

\end{document}